\documentclass[aps,prb,amsmath,amssymb,superscriptaddress,showpacs,floatfix,twocolumn]{revtex4-2}
\usepackage{graphicx,amsfonts,times,bm,amsmath,verbatim,color,array}
\usepackage{ifthen,braket,xcolor,bm}
\usepackage{braket}
\usepackage{cancel}                  
\usepackage{natbib}
\usepackage[colorlinks=true, allcolors=blue, breaklinks=true]{hyperref}
\usepackage[switch,columnwise]{lineno}

\graphicspath{{figures/}}

\begin{document}



\title{Universal scaling of quantum caustics in the dynamics of interacting particles}


\author{Monalisa Singh Roy}
\email{singhroy.monalisa@gmail.com,singhrm@biu.ac.il}
\affiliation{Department of Physics, Bar-Ilan University, Ramat Gan 52900, Israel}

\author{Jesse Mumford}
\affiliation{Homer L. Dodge Department of Physics and Astronomy, The University of Oklahoma, Norman, Oklahoma 73019, USA}
\affiliation{Center for Quantum Research and Technology, The University of Oklahoma, Norman, Oklahoma 73019, USA}
\author{D. H. J. O'Dell}
\affiliation{Department of Physics and Astronomy, McMaster University, 1280 Main Street W., Hamilton Ontario, Canada L8S 4M1}
\author{Emanuele G. Dalla Torre}
\email{emanuele.dalla-torre@biu.ac.il}
\affiliation{Department of Physics, Bar-Ilan University, Ramat Gan 52900, Israel}



\newcommand{\emanuele}[1]{{\bfseries\color{magenta} #1}}
\newcommand{\jesse}[1]{{\bfseries\color{orange} #1}}
\newcommand{\monalisa}[1]{{\bfseries\color{blue} #1}}

\date{\today}

\begin{abstract}
Recent theoretical studies have predicted the existence of caustics in many-body quantum dynamics, where they manifest as extended regions of enhanced probability density that obey temporal and spatial scaling relations. Focusing on the transverse-field Ising model, we investigate the dynamics initiated by a local quench in a spin chain, resulting in outward-propagating excitations that create a distinct caustic pattern. We calculate the scaling of the first two maxima of the interference fringes dressing the caustic, finding a universal  exponent of 2/3, associated with an Airy function catastrophe. We demonstrate that this property is universal in the entire paramagnetic phase of the model, and starts varying at the quantum phase transition (QPT).
This robust scaling persists even under perturbations that break the integrability of the model. We additionally explore the effect of boundary conditions and find that open boundaries introduce significant edge effects, leading to complex interference patterns. Despite these edge-induced dynamics, the overall power-law scaling exponent remains robust. These findings highlight the potential of quantum caustics as a powerful diagnostic tool for QPTs, demonstrating resilience against integrability-breaking perturbations and boundary condition variations.
\end{abstract}

\maketitle

\section{Introduction} \label{Introduction}

Caustics are regions of anomalously high amplitude in wave patterns where the classical ray approximation breaks down by predicting a diverging intensity. Well known everyday examples include the focusing of sunlight to form rainbows and the bright curved lines projected onto a table by a glass \cite{nye1999natural}.  In fact, caustics are a general feature of wave propagation in nonuniform backgrounds and occur across the full range of physical wave phenomena including gravitational lensing in optical and radio astronomy leading to Einstein rings \cite{Einstein1936,Nakamura1999,BARTELMANN2001,Berry_2021may}, as well as ship wakes \cite{Kelvin1905}, rogue waves \cite{ONORATO2013rogue}, and tsunamis \cite{Berry07} in oceanography. They have also been observed in quantum waves such as in cold neutron spectrometers \cite{KAWABATA1986,Cubitt2018}, electron microscope beams \cite{Petersen2013Jan}, and atomic Bose-Einstein condensates \cite{Huckans2009Oct,Rosenblum2014Mar,Mossman2021}. 

A key property of caustics is that they typically take on certain shapes which  are observed to consist of sharp lines that join at cusps. This is true whether they result from of a single focusing potential \cite{berry1980catastrophe} or from passage through random media \cite{Hohmann10,metzger2014,mathis2015caustics,Degueldre16}. An explanation for this behavior is provided by catastrophe theory \cite{Thom1975,Arnold1975} (CT) which predicts that there are certain special shapes of singularity that are stable against perturbations and hence occur generically without the need for fine tuning or perfect symmetry. This property is known as \textit{structural stability}. 
The special shapes, or `catastrophes', form a hierarchy, and in the two dimensional plane that will be most relevant to this work, there are only two: fold \textit{lines} that meet at cusp \textit{points}.  
At large scales, catastrophes appear singular, while at microscopic scales each type of catastrophe is smoothed out by a characteristic interference pattern. In the case of folds, the pattern is described by the Airy function, and in the case of cusps, by the Pearcey function \cite{Olver2010}. These patterns obey self-similar scaling of the their intensity and interference fringes characterized by scaling exponents known as Berry indices \cite{berry81}. 

Recent research has highlighted the importance of caustics in understanding the behavior of many-body quantum systems far from equilibrium. For instance, experiments \cite{Cheneau2012light} on the dynamics following a sudden quench in an interacting quantum many-body system realized using ultracold atoms in an optical lattice found that  the time evolution of the two-point correlation function displayed Airy function-like behavior \cite{Barmettler12}. Similarly, theoretical studies on the dynamics of the two and three-mode Bose Hubbard models and also the sine-Gordon model have shown that quenches in these systems generically lead to the development of caustics in Fock space \cite{ODell2012Oct,Mumford2017catastrophe,Mumford2019May,Kirkby2022Feb,Agarwal2023caustics} that are robust against decoherence \cite{Goldberg2019Dec}. These quantum caustics are regions of high probability in the space of many-body configurations and take on the shapes predicted by CT. The analogue of geometric rays are classical configurations, whereas the many body state is a superposition of multiple rays; on a caustic the density of rays diverges.  

In this paper we are interested in caustics that arise in the dynamics of quantum spin chains. In reference \cite{2019_Kirkby} it was shown that the light-cone-like structure that emerges following sudden quenches and governs the propagation of correlations along the chain as a function of space and time is in fact a caustic itself and agrees with the Lieb-Robinson bound \cite{Lieb1972}. For instance, in the transverse field Ising model (TFIM) the two edges of the light-cone are fold catastrophes and the Berry index $\zeta=2/3$ associated with the fold catastrophe gives the temporal scaling of the Airy function fringes that dress the cone as a function of the spin-spin coupling. Ref.~\cite{Riddell2023sep} found the same scaling for the light-cone appearing in the out-of-time-ordered correlator (OTOC) for a Heisenberg spin chain illustrating the universality of the CT predictions. 
Furthermore, it was found that adding weak integrability breaking terms to the Heisenberg spin Hamiltonian in the form of next-to-nearest-neighbor coupling only weakly perturbed the Airy fringes, confirming their structural stability. Applying strong integrability breaking that puts the system in the fully chaotic regime (obeying the eigenstate thermalization hypothesis \cite{DAlessio2016,Rigol2008,Depalma2015}) ultimately washes out the fringes. 
 However, since generic systems are neither fully integrable nor fully chaotic, but somewhere in between, the fact that caustics survive weak chaos and exhibit extreme amplitudes means they can still play a significant or even dominant role in the wave function and correlation functions in the general case, especially at short times.

 An important question that has not been analyzed in detail so far is the effect of quantum phase transitions (QPT) upon caustics. In reference \cite{2019_Kirkby} the density of vortex-anti vortex pairs that make up the finest scales of the interference pattern inside a  caustic was investigated and found to decrease sharply at the critical point in line with the expected critical slowing down of dynamics near QPTs. However, the general resilience of caustics to QPTs remains to be investigated. This is significant because the dynamics on either side of the transition can be due to very different types of excitations (quasiparticles).  Understanding this resilience is crucial for assessing the robustness of caustics as diagnostic tools for QPTs and other critical phenomena in many-body quantum systems.

In the present work, we investigate the dynamics of the  TFIM, a paradigmatic interacting many-body system, as well as an integrability breaking extension, to explore the formation and scaling behavior of quantum caustics. By quenching the central spin in a chain that is initially in a product state, we generate an outward-propagating domain light-cone that creates a distinct quantum caustic pattern. 
We focus on the scaling behavior of the first two maxima (fringes) in this pattern, identifying a universal scaling exponent before the QPT 
and examining how this scaling is affected by integrability-breaking 
interactions and varying boundary conditions.

Our findings demonstrate that the scaling behavior of quantum caustics is robust against small integrability-breaking perturbations until a critical threshold is reached, beyond which the caustic pattern degrades. Additionally, we show that while periodic boundaries maintain the translational invariance and uniform propagation of domain walls, open boundaries introduce significant edge effects that lead to complex interference patterns. Despite these edge-induced dynamics, the overall power-law scaling exponent remains robust, underscoring the potential of quantum caustics as a reliable diagnostic tool in many-body quantum systems.
Through this study, we aim to advance the understanding of quantum caustics and their application in identifying QPTs, providing a foundation for further exploration of caustic phenomena in complex quantum systems.

The rest of this paper is organized as follows. In Section \ref{Model and Method} we outline the model and main observable (relative fringe width) that we use for our study of caustics on spin chains. In Section \ref{sec:weak} we study the weak-coupling limit where the low-lying excitations are spin flips so that  we can initiate the dynamics from a single spin flip and propagate it in time analytically in a straightforward manner. We then map the resulting light cone onto an Airy function.  The theoretical aspects of this mapping are further discussed in Section \ref{sec:theory} where we show that the Airy function solution is not limited to  weak coupling. In the following sections we numerically study the roles of integrable (Section \ref{sec:numerics}) and nonintegrable (\ref{sec:nonintegrable}) perturbations and boundary conditions (Section \ref{sec:boundary}). In Section \ref{sec:summary} we summarize our results and give our conclusions.


\section{Quench dynamics of the TFIM} \label{Model and Method}

We consider a one dimensional (1D) spin model with nearest neighbour Ising coupling $J^{xx}$ and transverse field $h^{z}$. This model, often referred to as the transverse-field Ising model (TFIM), corresponds to the Hamiltonian
\begin{equation} \label{eq.1: Hamiltonian}
 H = J^{xx} \sum_{j=1}^N S_j^{x} S^{x}_{j+1} + h^{z} \sum_{j=1}^N S^{z}_j.
\end{equation}
Here, $N$ is the number of spins, which we assume to be odd; $S_j^{x}$ and $S_j^{z}$ are, respectively, the spin-1/2 operators in the $x-$ and $z-$ directions at position $j$. The TFIM model is a paradigmatic example of an integrable model showing a quantum phase transition (QPT) at $h_z=J^{xx}$ and has been used widely to study nonequilibrium effects such as prethermalization \cite{marcuzzi2013prethermalization}, dynamical phase transitions \cite{heyl2013dynamical}, and more. 


{Throughout this paper, we consider the following quantum quench protocol:} We initialize the system in a product state where all spins point down, except for the middle one, at $j_0=(N+1)/2$, which points up
\begin{align} \label{eq_2_psi0}
|\psi(t=0)\rangle = \left(\prod_{j<j_0}|\downarrow\rangle_j\right) |\uparrow\rangle_{j_0} \left(\prod_{j>j_0}|\downarrow\rangle_j\right)
\end{align}
This initial state has can be easily prepared in quantum simulators with single-site addressability, and has been realized experimentally with both ultracold atoms \cite{fukuhara2013quantum} and trapped ions \cite{Jurcevic2014}. {Under the TFIM, the dynamics} creates an outward propagating light-cone, starting from $j=j_0$.
{The spins residing outside the light-cone are only very weakly perturbed, in accordance with the Lieb-Robinson bound, which describes an exponential suppression of spatial correlations in spacelike separated regions. In this work, we focus on the interference patterns inside the light-cone and associate them to quantum caustic fringes with a robust scaling.}

To observe the scaling properties of the interference pattern, we consider the time evolution of the $z$ component of the $j-$th spin, $\langle S^z_j(t)\rangle$. This function is non-monotonous and is characterized by an exponential growth, followed by decaying oscillations. We denote the time location of the first two peaks of these oscillations by $t_{1}$ and $t_{2}$, respectively, see Fig.~\ref{fig:schematic}(b). We, then, define a unitless quantity which will serve as our ``order parameter'': the normalized time delay between the first two peaks,
\begin{equation}
O_{1,2}(j) = (t_2 - t_1)/{t_1}.\label{eq:O12}
\end{equation}
As we will show, the function $O_{1,2}(j)$ follows a power-law behavior with a universal exponent 2/3.
{Importantly, the universal scaling is a many-body property that emerges at large distances}
from the quenched spin, i.e. for $|j-j_0|\gg 1$.  
We will show that this result remains true even in the integrability broken limit and represents a universal feature for identifying QPTs in such many-body systems. 


\section{Weak coupling limit}
\label{sec:weak}

We open our analysis by considering the weak coupling limit $J^{xx}\ll h_z$. In this limit, the ground state of the TFIM is well approximated by $\prod_j \ket{\downarrow}_j$ and the low-lying excitations are spin flips with a well-defined momentum $k$, which we denote by $\ket{k}$. Up to first order in perturbation theory, their energy with respect to the ground state is given by $\epsilon_k = h^z + J^{xx} \cos(k)$. The spread of correlations in the system is governed by the Lieb-Robinson bound with velocity $v_{\text{\tiny{LR}}}=\max_k \left\vert {\mathrm{d} \epsilon_k}/{\mathrm{d} k} \right\vert = J^{xx}$. Hence, an excitation initially positioned at $j=j_0$ will lead to a light-cone delimited by
\begin{align}
|j-j_0| = J^{xx} t \label{eq:lightcone}.
\end{align}

The initial state chosen for our calculation corresponds to a single spin flip at position $j_0$ and can be written as a linear combination of $\ket{k}$ modes,
\begin{align}
\ket{\psi(t=0)} =|j_0\rangle \equiv \frac{1}N\sum_k e^{ik j_0}\ket{k}.
\end{align}
Here, the sum over $k$ is performed over the set $k=(2\pi/N)\times (0,1,...,N-1)$ and we assumed periodic boundary conditions. Accordingly, the magnetization at position $j$ can be expressed as
\begin{align}
    \langle S^z_j(t)\rangle = -\frac12+\left|\langle j|\psi(t)\rangle \right|^2  \ , \label{eq:Szjt}
\end{align}
which depends on the probability of a spin flip having occurred at that position.
Here, $|\psi(t)\rangle$ is the time evolution of $\ket{\psi(t=0)}$ and is given by
\begin{align}
\ket{\psi(t)}  &= \frac1N\sum_k e^{i k j_0 - i \epsilon_k t}|k\rangle \\
&= \frac1N{e^{-i h^z t}}\sum_k e^{i k j_0 - iJ^{xx} \cos(k) t}\ket{k},\label{eq:psit}
\end{align}
where we have set $\hbar=1$. By combining Eqs.~(\ref{eq:Szjt}) and (\ref{eq:psit}) we find that 
\begin{align}
\langle S^z_j(t)\rangle 
& =-\frac12+ \left|\frac1N\sum_k e^{i\Phi_0(j,t,k)}\right|^2\label{eq:Szj},
\end{align}
with
\begin{align}
\Phi_0(j,t,k)= k(j-j_0)-J^{xx} \cos(k) t.
\end{align}




\begin{figure}[t]

\includegraphics[width=0.5\textwidth,trim={0.0cm 1.45cm 1cm 2.5cm},clip]{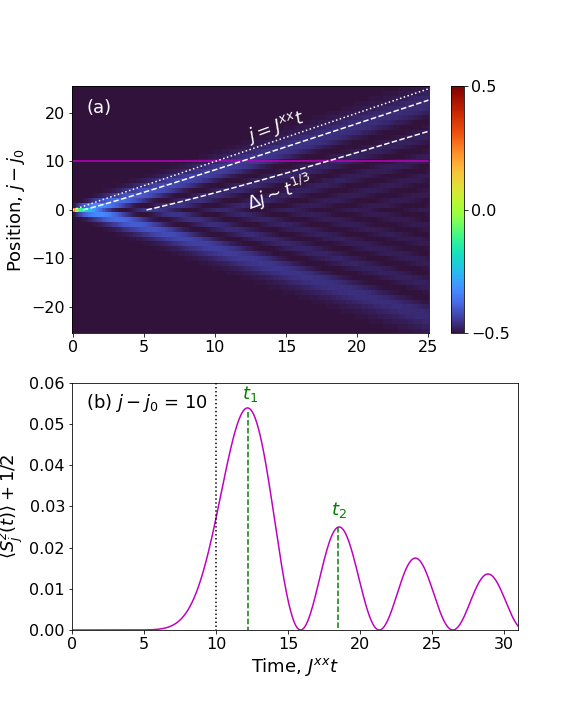}
\caption{Analytical solution of the TFIM in the in the continuum and weak-coupling ($J^{xx}\ll h^z$) limits. (a) Local magnetization as a function of position and time, $\langle S^z_j(t)\rangle $, Eq.~(\ref{eq:airy2}). The dotted line highlights the position of the edge of the light-cone, which corresponds to the origin of the Airy function. The dashed curves highlight the position of the first two peaks, denoted by $t_1$ and $t_2$. (b) Horizontal cut, using the same conventions for the dashed and dotted curves.}
\label{fig:schematic}
\end{figure}

To describe the scaling properties of the interference patterns in the vicinity of the light-cone, we can perform a Taylor expansion of $\Phi_0$ in its vicinity.  Defining $\Delta j = |j - j_0| -J^{xx}t$ 
and expanding to third order in $k-k_0$ with $k_0=\pi/2$, we obtain
\begin{align}
\Phi_0(j,t,k)\approx k_0 j + (k-k_0) \Delta j +\frac13J^{xx}(k-k_0)^3 t .\label{eq:phi0}
\end{align}
Substituting this result into Eq.~(\ref{eq:Szj}) we find
\begin{align}
S_j^z(t) \approx -\frac 12 +\left|\frac{1}N\sum_k e^{i(k-k_0)\Delta j+i\frac13J^{xx}(k-k_0)^3 t}\right|^2.
\end{align}

In the continuum limit we can transform the sum over to an integral, $\sum_k \to (N/2\pi)\int_{-\infty}^\infty dk$, and introducing $s=(k-k_0)t^{1/3}$, obtain
\begin{align}
S_j^z(t) &\approx-\frac12+ \frac {1}{(2\pi)^2t^{2/3}}\left| \int_{-\infty}^\infty ds~ e^{iJ^{xx}(C(j,t)s+\frac13s^3)}\right|^2,\label{eq:airy}
\end{align}
where we defined $C(j,t)=\Delta j/(J^{xx}t^{1/3})$ and $\Delta j$ is the distance from the light-cone, defined above Eq.~(\ref{eq:phi0}). By rescaling $s\to s/(J^{xx})^{1/3}$, we can rewrite this expression in terms of the square of the Airy function ${\rm Ai}$ as
\begin{align}
S_j^z(t) \approx -\frac12+\frac{1}{(J^{xx}t)^{2/3}}{\rm Ai^2}\left[\frac{\Delta j}{(J^{xx}t)^{1/3}}\right],\label{eq:airy2}
\end{align}
Note that in this weak coupling limit, the dynamics is a function of the unitless parameter $J^{xx}t$ only. The resulting interference pattern is shown in Fig~\ref{fig:schematic}.

Equation (\ref{eq:airy}) demonstrates an explicit relation between the time evolution of a physical observable, namely the magnetization following a quantum quench, and caustics. As mentioned above, our quantity of interest is the time delay between two subsequent peaks, $\Delta t = t_2-t_1$, which corresponds to a fixed distance between values of the argument of the Airy function, or $\Delta j \sim \Delta t \sim t^{1/3}$. Close to the edges of the light-cone $t\sim |j-j_0|$, leading to 
\begin{align} O_{1,2}(j) \sim |j-j_0|^{-2/3},\label{eq:scaling}
\end{align}
which highlights the dependence of our order parameter on the Berry index $2/3$, characteristic of the Airy function.


\section{Coalescing saddles and catastrophe theory}
\label{sec:theory}


Before we present the numerical results of the paper, we will highlight exactly how the analysis in the previous section falls under the wider purview of CT.  One of the main results of CT is that near coalescing saddles, the phase function can be mapped to a  universal form defined in terms of the number of control parameters $\textbf{C} = \{C_1, C_2, ... \}$ and the number of state variables $\textbf{s} = \{s_1, s_2, ... \}$.   In the simplest case where there is a single state variable, the universal form---or generating function---is a polynomial in the state variables with coefficients given by the control parameters \cite{Olver2010}
\begin{equation}
\Phi_Q(s;\textbf{C}) = \frac{s^{Q+2}}{Q+2} + \sum_{n = 1}^Q \frac{C_n s^n}{n}.
\label{eq:GF}
\end{equation}
where $Q$ is the maximum number of coalescing saddles.  The condition for saddles to exist is ${\partial \Phi_Q}/{\partial s} = 0$ and when they coalesce, the generating function is stationary to higher order ${\partial^2 \Phi_Q}/{\partial s^2} = 0$.  Another result from CT is that $\Phi_Q(\textbf{s},\textbf{C})$ is the most general way to capture the behaviour of coalescing saddles.  This in turn means that catastrophes are stable to perturbations. For example, one may worry that higher order terms in the Taylor expansion in Eq.\ (\ref{eq:phi0}) have been neglected and that the resulting wave function is not an Airy function. This is not the case: even if the parameters in our cubic generating function are approximate, CT guarantees that the true generating function near a fold catastrophe is a cubic and hence an Airy function is the exact local wave function.  

\begin{figure*}[t]
\centering
\includegraphics[width=0.98\linewidth]{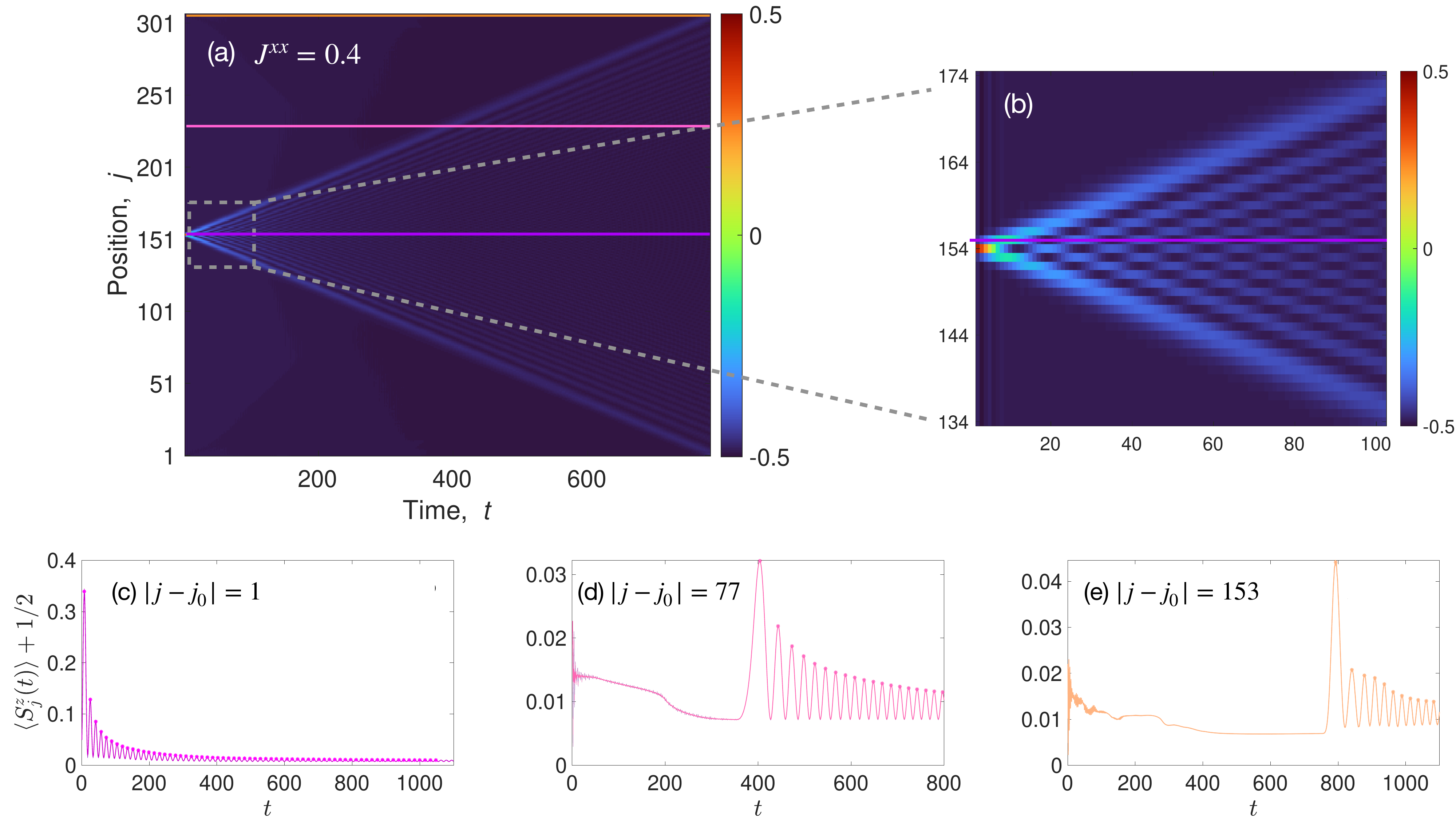}
\caption{Numerical simulation of the TFIM with $N=307$ spins for $J^{xx}=0.4$. (a) Local magnetization as a function of position and time, $\langle S^z_j(t)\rangle$. The interference pattern gives rise to a light-cone that emanates from the quenched spin at $j_0=154$. (b) Zoom-in: a clear visualization of the caustics. (c-e) Local magnetization as a function time at fixed position, $j$. These curves are horizontal cuts of (a) at the positions denoted by the colored lines. Each curve is characterized by a pronounced peak whose time delay grows linearly with $|j-j_0|$ as predicted by the Lieb-Robinson bound. At later times the system shows decaying oscillations with a universal scaling.}\label{fig2new}
\end{figure*}


CT becomes applicable in physics when a variational approach is needed such as in the case of finding the ground state by minimizing the energy.  In that case, $\Phi_Q$ is the energy, $\textbf{s}$ give the state of the system and $\textbf{C}$ are  external fields that can be tuned.  However, when it comes to catastrophes in the dynamics, which is the focus of this paper, $\Phi_Q$ assumes the role of an \textit{action}. In this case, the condition for saddles to exist corresponds to extremizing the action to determine classical trajectories, while the coalescing condition determines the focusing of those trajectories. This relationship establishes a link to quantum mechanics through the Feynman path integral where the control parameters $\textbf{C}$ are functions of time and space and the state variables $\textbf{s}$ label each path.  In this context, the diffraction catastrophes of the form

\begin{equation}
\Psi_Q(\textbf{C}) \propto J^{n/2}\int_{-\infty}^{\infty}...\int_{\infty}^{\infty} d^nse^{iJ \Phi_Q(\textbf{s};\textbf{C})}    
\label{eq:WFQ}
\end{equation}
are useful in describing the focusing of quantum wave functions.  In the simplest case where $Q=1$ we have the Airy function \cite{2019_Kirkby}
\begin{equation}
\Psi_1(C) \propto J^{1/6} \mathrm{Ai}(J^{2/3} C)
\label{eq:scale}
\end{equation}
which is utilized to describe the coalescence of two saddles.  Equation \eqref{eq:scale} exhibits the wavenumber scaling exponents which are unique to each diffraction catastrophe. Specifically, the Berry exponent of $2/3$ governs the scaling of the interference fringes in the wave wave function.

Comparing this general framework to the quenched TFIM previously discussed, we can consider $\Phi_0$ in Eq.~\eqref{eq:phi0} as a generating function. Being a cubic function of $k$, it has two saddle points (classical solutions) which can coalesce as the parameters $\Delta j$, $J^{xx}$ and $t$ are varied. The locus of spacetime points where this coalescence takes place defines the location of the caustic and can be found by demanding both the first and second derivatives to vanish, $d\Phi_0/dk=d^2\Phi_0/dk^2=0$. The latter condition is satisfied for $k=\pm\pi/2$, leading to $ \Delta j = \pm J^{xx} t$.  This is precisely the equation for the edges of the light-cone calculated using the Lieb-Robinson bound in Eq.~(\ref{eq:lightcone}). Hence, light-cones can be considered as caustics described by CT.  Therefore, it should come as no surprise that the polarization $S^z_j(t)$ in the vicinity of the edges of the light-cone takes the form of a squared Airy function since it is the simplest diffraction catastrophe as shown in Eq.~\eqref{eq:scale}. As explained Ref.~\cite{2019_Kirkby}, this analysis is not limited to the weak coupling limit: for arbitrary values of the coupling $J^{xx}$, when a single quasiparticle (Bogoliubov fermion) is excited, the resulting dynamics is described by Airy functions. 

What remains an open question is the form of light-cones when integrability is broken and whether scaling exponents similar to the Berry exponents can be extracted from their intensity and interference patterns.  Some progress has been made along these lines with an analysis of how the amplitude of the light-cone scales with respect to the distance from its highest peak in out-of-time-order correlators (OTOCs) in the Heisenberg spin model \cite{Riddell2023sep}.  This analysis demonstrated that the scaling behavior differs significantly if the system is deep within the nonintegrable phase. However, it is also showed that the Airy function dressing the light cone wavefront is not perturbed at first order by integrability breaking, i.e.\ quantum caustics have structural stability against chaos too (this situation is reminiscent of the Kolmogorov-Arnold-Moser theorem guaranteeing the survival of some tori in the phase space of weakly chaotic systems \cite{Brandino2015}), and hence the scaling is preserved in the more generic weakly chaotic regime.  Next, we explore the scaling of the interference fringes within the light-cone in the presence of a QPT and integrability breaking.

\section{Numerical studies of the TFIM and the role of the QPT}
\label{sec:numerics}


To study the stability of the caustics' scaling beyond the weak coupling limit, we numerically simulate the evolution of a spin chain under the TFIM Hamiltonian, Eq.~\eqref{eq.1: Hamiltonian}. To reach appreciable system sizes, we apply a matrix-product-state (MPS) protocol \cite{wall2012out}, implemented using in the itensor package~\cite{2020_iTensor}. In our calculations we prepare the initial state $|\psi(t=0)\rangle$ defined in Eq.~\eqref{eq_2_psi0} by flipping a single spin in the middle of the chain. The evolution is then performed according to the first-order Trotter expansion of the Hamiltonian with time step $\Delta t$, i.e. by alternating layers of $\exp(-iJ^{xx}\Delta t\sum_j S_j^xS_{i+1}^x)$ and $\exp(-ih^z \Delta t\sum_j S_j^z)$ operators. Without loss of generality we set $h^z$ to one and check that $\Delta t$ is small enough to ensure the validity of the Trotter expansion. In practice $\Delta t =1$ is small enough for most purposes. Importantly, this description of the time evolution can be directly implemented on state-of-the-art quantum computers, as demonstrated in several recent experiments, see for example, Refs.~\cite{2019_Arute,2020_kjaergaard,2021_Azses, 2021_Kyprianidis, 2022_Xiao,2024_Fauseweh, 2024_miessen}.

{
}
\begin{figure*}
\centering

\begin{tabular}{lcr}    
 
        \includegraphics[height=0.35\linewidth]{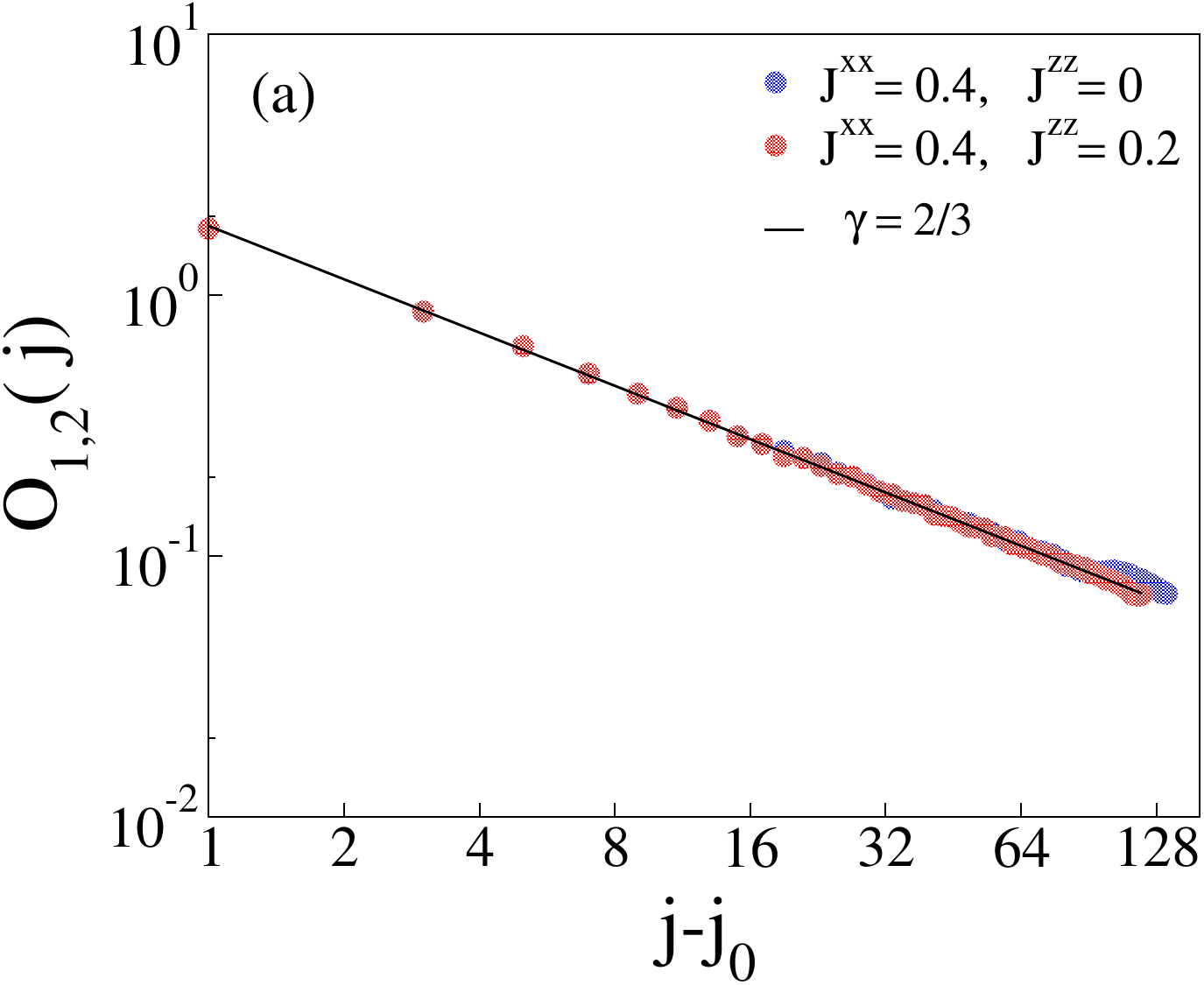}
& \hfill
&
        \includegraphics[height=0.35\linewidth]{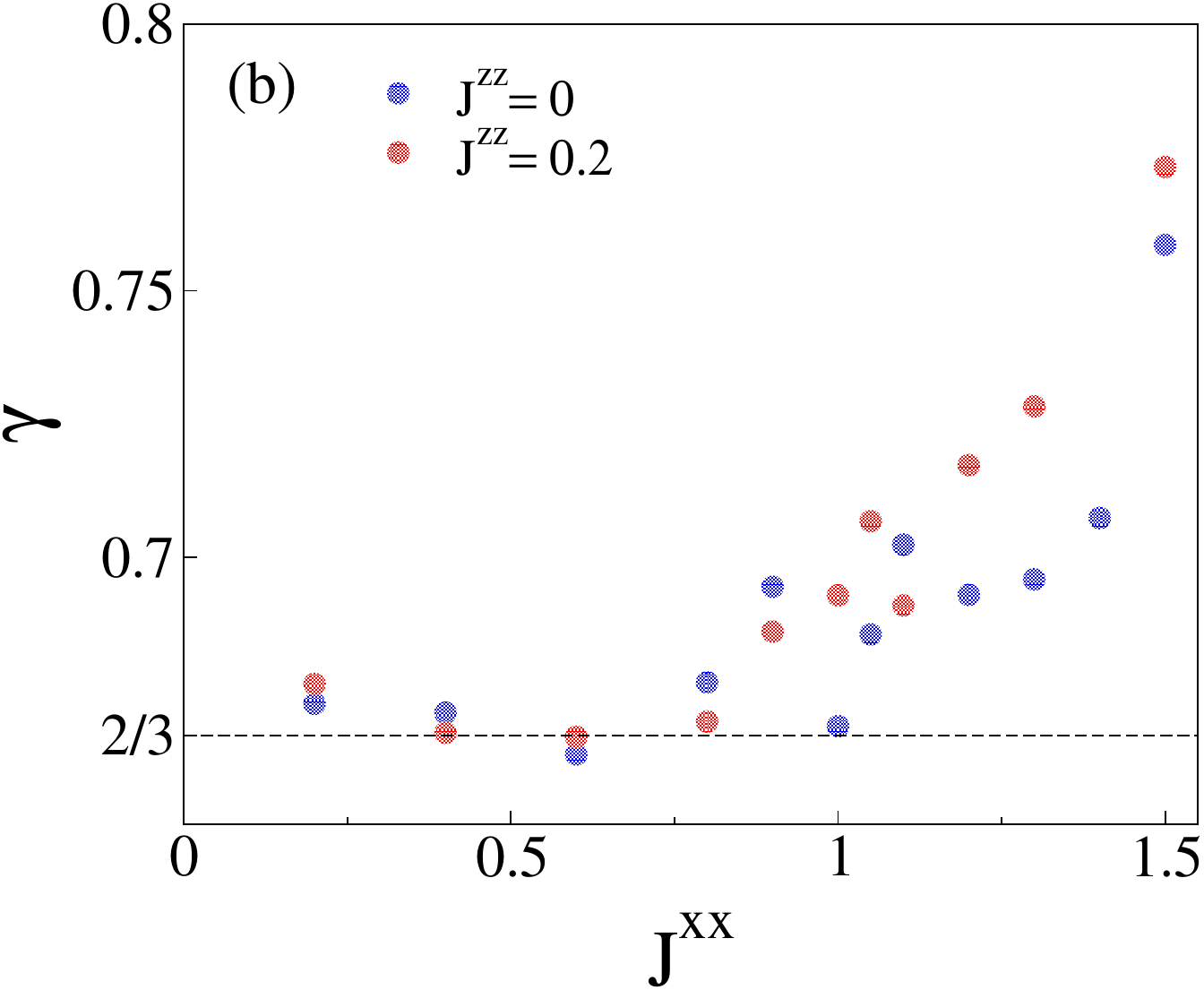}
\end{tabular}
\caption{Scaling analysis of the numerical simulations of the TFIM with $N=307$ spins. (a) Normalized time delay between the first two peaks, $O_{1,2}(j)$, as a function of the distance from the quenched spin, $j - j_0$. The straight line corresponds to a power-law fit $O_{1,2}(j) \sim (j-j_0)^\gamma$. (b) Fitted power-law exponent $\gamma$ as a function of the Ising coupling $J^{xx}$, with and without the integrability breaking term $J^{zz}$. The universal scaling exponent of $2/3$ is consistent with the theory of quantum caustics, remaining robust until the phase transition point, $J^{xx}/h^z = 1$.}\label{fig3new}

\end{figure*}

\begin{figure*}
\centering
\includegraphics[width=0.48\linewidth]{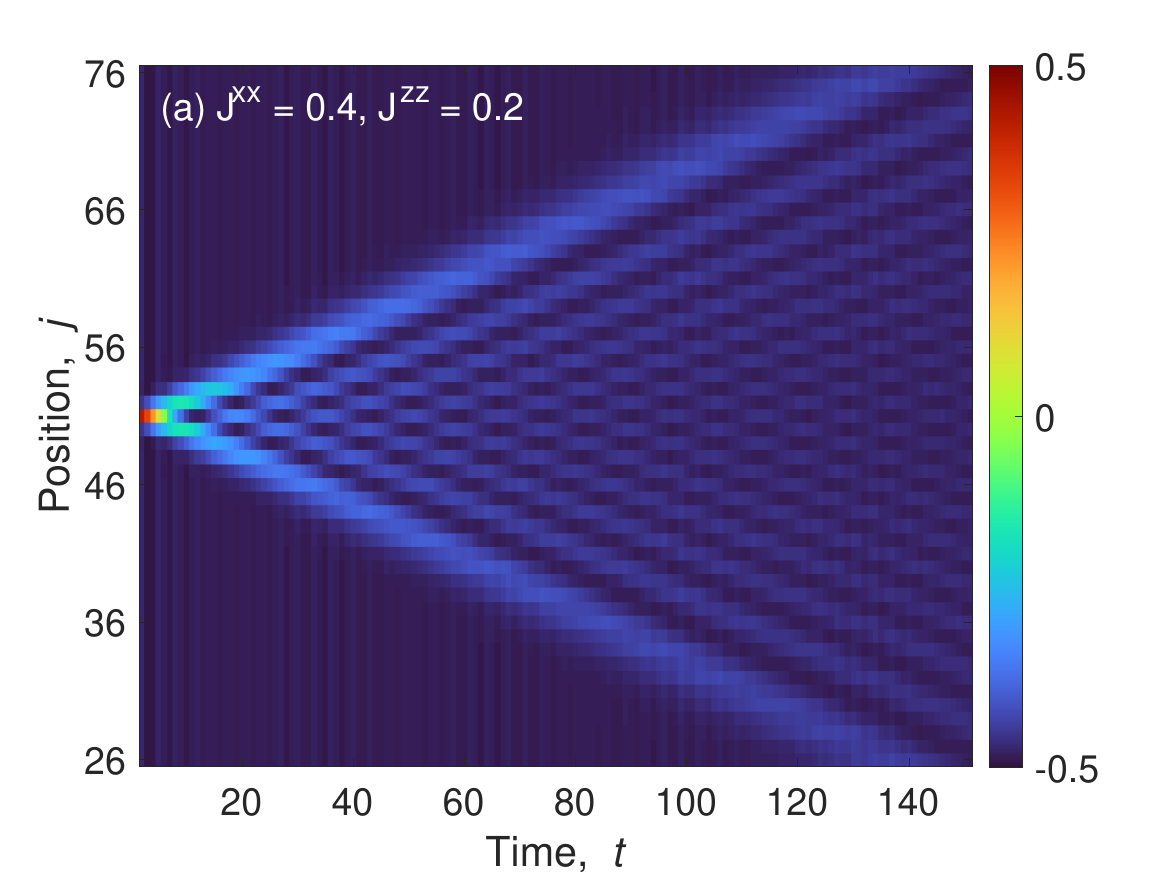}
\includegraphics[width=0.48\linewidth]{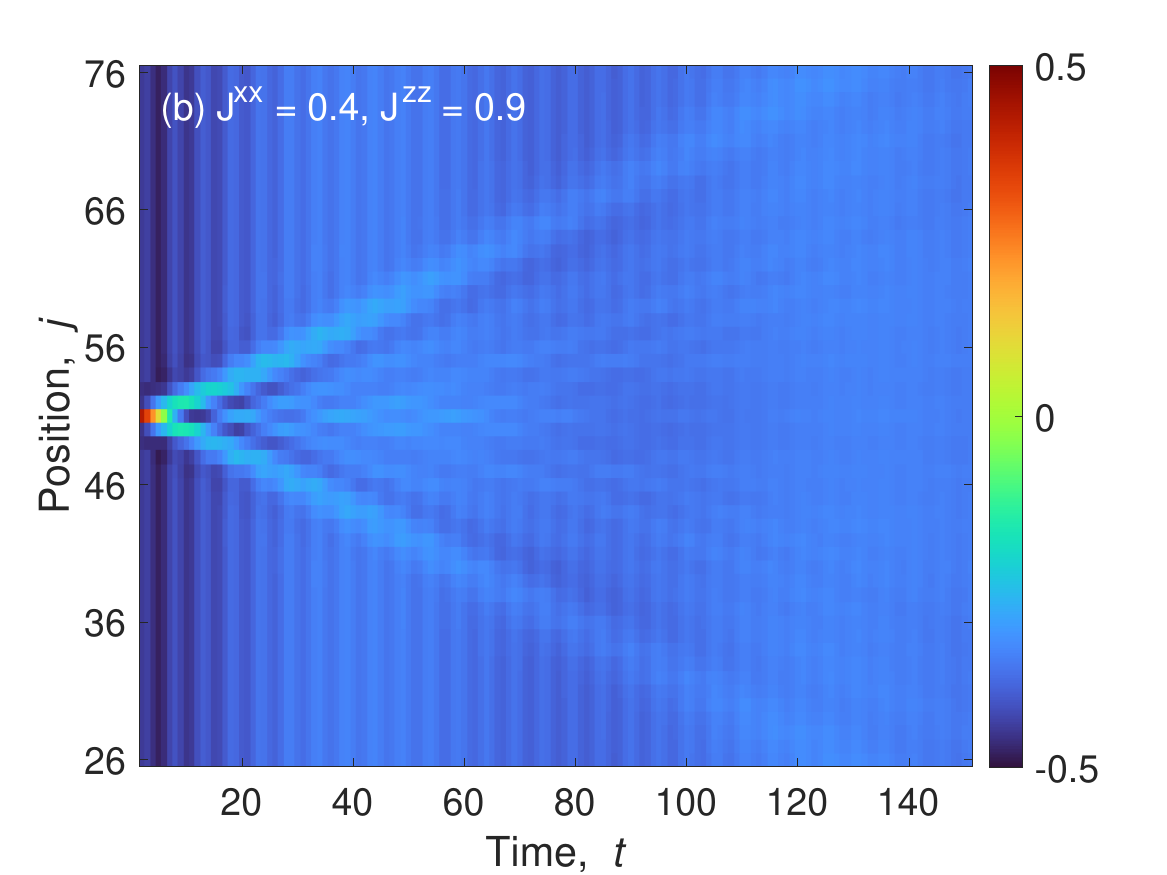}

\caption{Numerical simulations of the TFIM for two different values of the integrability-breaking parameter $J^{zz}$ (with $N=101$). The caustic patterns are resilient to intermediate values of this coupling, and fade out at large values of $J^{zz}$.
} \label{fig4new}
\end{figure*}

Let us first consider the limit of $J^{xx} \ll h^{z}$, where the system can be well described by a single-particle picture. In this regime, we expect to find a well-defined quantum caustic pattern, characterized by a
universal scaling with a power-law coefficient of $-2/3$. 
The expected pattern can indeed be clearly seen in Fig.~\ref{fig2new}(b).  Subfigures \ref{fig2new}(c), (d),  and (e) show the evolution of the local magnetization $S_j^z(t)$ at different positions.  It can be observed that the speed of propagation adheres to the Lieb-Robinson limit. Consequently, the time needed to reach the first maximum, $t_1$ grows linearly with $\Delta j = j - j_0$, where $j_0$ is the location of the quenched spin at the center. By tracking the first and second maxima and extracting $t_1$ and $t_2$ for each spin, we can compute $O_{1,2}(j)$, defined in Eq.~(\ref{eq:O12}). This function is shown in Fig.~\ref{fig3new} for $J^{xx}=0.4$ and demonstrates the expected power-law behavior. 

By repeating the same calculation for different values of $J^{xx}$, we find that the scaling exponent of 2/3 is unvaried for the entire paramagnetic phase of the TFIM, i.e. for all $J^{xx}<h_z$. For larger values of the interaction, we observe a monotonous increase of this parameter. The reason for this change is related to the initial state that has been chosen for our calculation. For $J^{xx}<h_z$ this state is adiabatically connected to the ground state of the Hamiltonian and the shape of the light-cone can approximated by the spreading of a single excitation. In the opposite limit, this approximation breaks down. Because these two regimes are equivalent up to a dual transformation, the results shown so far are valid in the opposite regime, for the appropriate initial state. To observe a universal scaling we should have begun from a state that is adiabatically connected to a local excitation, namely a state polarized in the $x-$ direction with a single domain wall at $j=j_0$.

\begin{figure*}[t]
\centering
\includegraphics[width=0.48\linewidth]{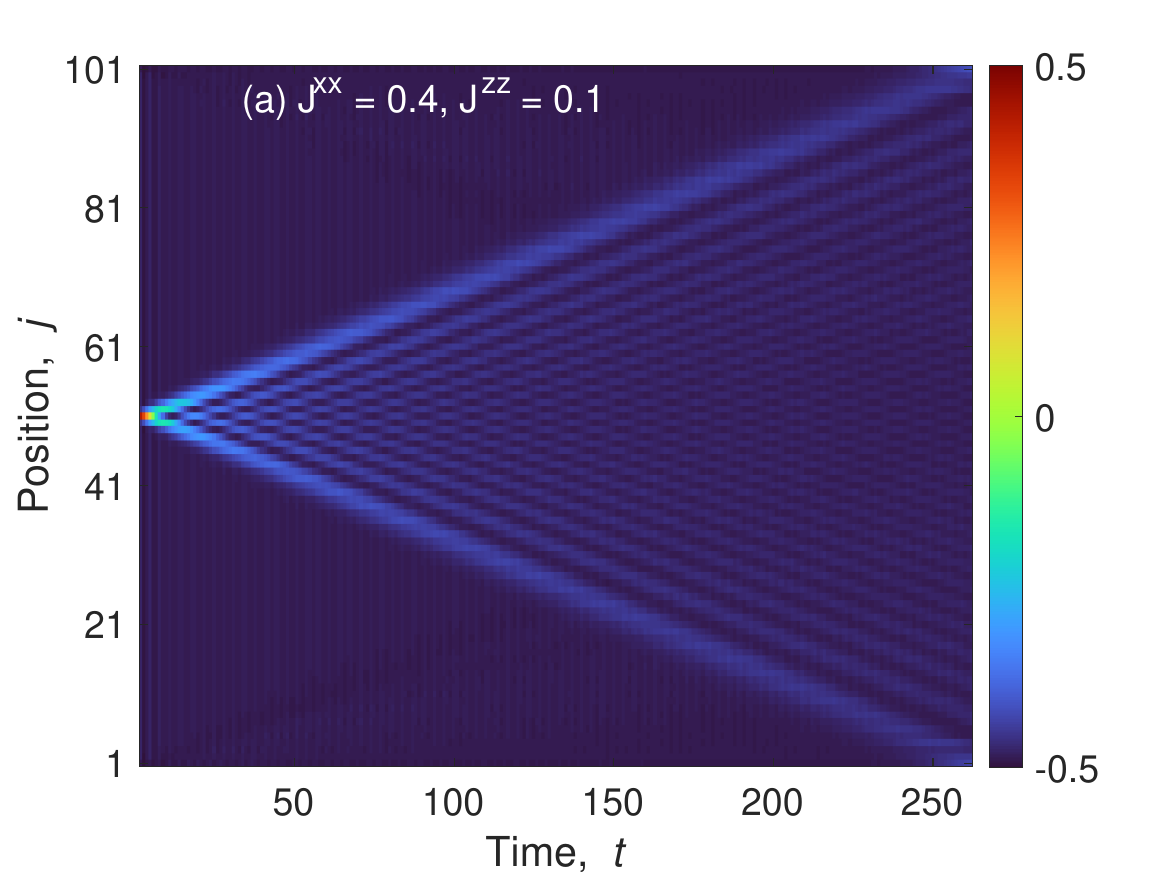}
\includegraphics[width=0.48\linewidth]{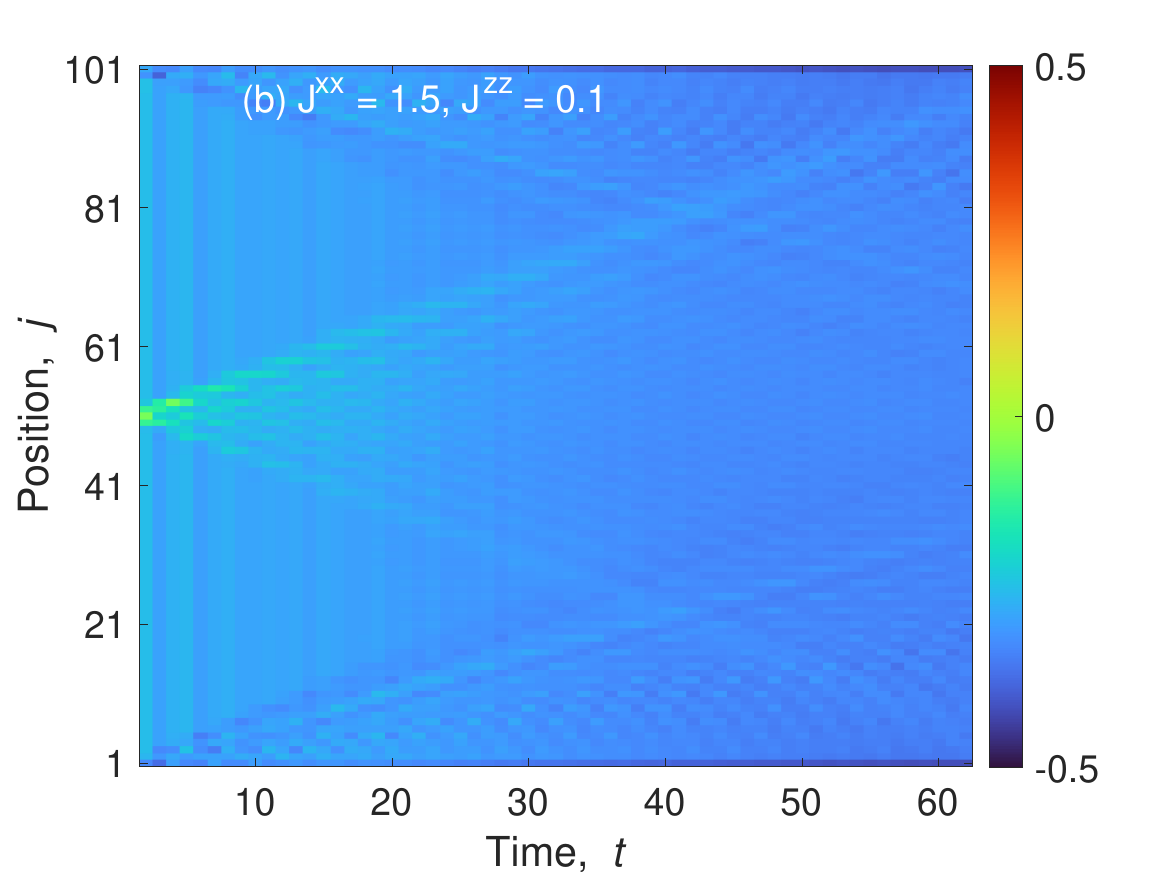}
\caption{Numerical simulations of the TFIM  with open boundary conditions (with $N=101$). (a) For $J^{zz}<h^x$, the interference pattern is the same as periodic boundary conditions. (b) For $h_x>J^{xx}$, the system develops pronounced inward propagating light-cones that emanate from the edges.}
\label{fig5new}
\end{figure*}

\section{Role of integrability breaking terms}
\label{sec:nonintegrable}

The quantum caustics observed in the system are expected to be robust and exhibit universal scaling. To test this robustness, we study the effects of integrability breaking by introducing a $J^{zz}$ interaction term. The Hamiltonian of the system is now described by 
\begin{equation} \label{eq.2: Int_broken_Hamiltonian}
    {H} = \sum_{\langle i, j \rangle} \left( J^{xx} {S}^x_j {S}^x_j + J^{zz} {S}^z_j {S}^z_j \right) + h_z\sum_j {S}^z_j.
\end{equation}

We investigate how the addition of this term affects the quantum caustics and the scaling exponent of the order parameter. We find that for small values of the integrability breaking parameter $J^{zz}$, the quantum caustics and the scaling exponent for the order parameter remain unchanged, see Figs. \ref{fig3new} and \ref{fig4new}. This indicates that the caustic pattern is robust to weak integrability breaking, preserving the universal scaling behavior.  As $J^{zz}$ increases, the caustic pattern begins to fade and the magnetization becomes more homogeneous throughout the space-time plane. At higher values of $J^{zz}$  the pattern is eventually lost. This loss of the caustic pattern at higher $J^{zz}$ values signifies a transition away from the universal scaling behavior observed in the integrable limit.


This investigation into the integrability breaking limit of the TFIM highlights the delicate balance between integrability and quantum chaos. The robustness of the quantum caustics at small $J^{zz}$ values underscores the potential for these patterns to persist in near-integrable systems, while the loss of caustics at higher $J^{zz}$ values marks a clear boundary where integrability is sufficiently broken to alter the system's dynamical properties fundamentally. These findings suggest further avenues for exploring the interplay between integrability, chaos, and universal scaling in quantum many-body systems, with potential implications for understanding quantum phase transitions and the robustness of emergent phenomena in complex quantum systems.

\section{Effect of the boundary conditions}
\label{sec:boundary}

Thus far, all calculations have been presented for periodic boundary conditions. The main results, including the power-law scaling exponent 2/3, do not depend on this particular boundary condition. Under periodic boundary conditions, the system is translationally invariant, and the light-cone created by quenching the central spin propagates uniformly outward without encountering any boundaries. This setup ensures that the observed quantum caustics and the associated power-law scaling are not influenced by the system's finite size or edge effects.

Fig.~\ref{fig5new} illustrates the impact of open boundary conditions on the local magnetization $S^z_{j}(t)$. Here, we observe that the light-cones originating from the edges of the system travel inward, creating additional features in the magnetization profile. These edge-induced light-cones can interact with those propagating from the central quench, leading to enhanced or diminished local magnetization depending on the relative phases of the propagating fronts. 
For $J^{xx} > h_{z}$, the system exhibits more pronounced boundary-induced features. The light-cones from the edges and the center can merge or reflect off each other, leading to richer dynamical behavior. This regime highlights the sensitivity of the system's dynamics to boundary conditions, which can be crucial for experimental realizations where open boundaries are more common.




\section{Summary and  Conclusion} \label{sec:summary}

In this study, we introduce a novel tool for identifying quantum phase transitions (QPT) in many-body models through the observation of quantum caustics. Using the transverse field Ising model (TFIM) as a case study, we demonstrate how the spreading of quantum caustics created by the two outward-propagating edges of a light-cone can help identify the nature of QPTs by probing its long wavelength quasiparticles. Our analysis encompasses both integrable and non-integrable limits of the model, revealing the robustness of the caustic pattern and the associated scaling behavior across different regimes.

We initiate our investigation with the 1D TFIM, where the Hamiltonian is defined by interactions mediated by $J^{xx}$ in the presence of a transverse field $h_{z}$. In the paramagnetic phase, for $J^{xx}<h_z$ the ground state is polarized downwards. By quenching the central spin in the opposite direction, we induce the formation of a light-cone that propagates outward, creating a distinct quantum caustic pattern observable in the local magnetization $S^z_{j}(t)$. Our results show that the time delay of these caustics follows a universal power-law behavior with an exponent of $-2/3$. This effect is robust in the entire paramagnetic phase, for $J^{xx} < h_{z}$. Beyond the QPT transition, the scaling factor increases monotonically with $J^{xx}$, indicating a shift in the system's dynamical properties.

The robustness of this scaling behavior is further tested by introducing $J^{zz}$ interactions, breaking the integrability of the model. We find that the quantum caustics and the associated scaling exponent remain unchanged for small  integrability breaking.  However, as $J^{zz}$  increases, the caustic pattern degrades and eventually disappears, hindering the observation of the universal scaling behavior observed in the integrable limit. This highlights the delicate balance between integrability and quantum chaos in determining the system's dynamics.

Additionally, we explore the effects of boundary conditions on the TFIM. While periodic boundary conditions are characterized by a single light-cone that emanates from the quenched spin, open boundary conditions introduce significant edge effects, particularly when $J^{xx} / h_{z} > 1$. These edge-induced light-cones interact with those from the central quench, leading to complex interference patterns and modified local dynamics. 

Our findings emphasize the potential of quantum caustics as a diagnostic tool for QPTs in many-body systems. The observed robustness of the caustics pattern and its scaling behavior against integrability breaking and boundary conditions suggests a broader applicability of this approach in studying complex quantum dynamics. Future studies could extend this analysis to other boundary conditions, such as mixed or time-dependent boundaries, to further elucidate their impact on quantum caustics and scaling behaviors. These insights will aid in designing more precise experiments and simulations to probe the fundamental aspects of quantum many-body dynamics.

By leveraging the universal scaling properties of quantum caustics, we open new avenues for exploring and understanding quantum phase transitions and the resilience of emergent phenomena in complex quantum systems. This approach not only enhances our theoretical understanding but also provides practical tools for experimental realizations in quantum computers and ultracold atom experiments.

\acknowledgements{This research was supported in part by the International Centre for Theoretical Sciences (ICTS) by the participation of DHJO and EGDT in the program Periodically and quasi-periodically driven complex systems  (code: ICTS/pdcs2023/6). DHJO acknowledges support from the Natural Sciences and Engineering Research
Council of Canada (NSERC) [Ref. No. RGPIN-2017-06605]. MS and EGDT are supported by the Israel Science Foundation, grant No 154/19 and 151/19.}




\bibliography{Quantum_Caustics} \label{References}

\appendix




\end{document}